\documentclass[a4paper,12pt]{article}

\usepackage{graphicx}
\usepackage{caption}
\usepackage[subrefformat=parens]{subcaption}

\begin{document}

%\vspace{-5pt}
\begin{center}
{\bf \Large A relation between track length and deposited energy in a homogeneous calorimeter by Geant4 simulation at high energy
 }\\[14pt]

{\large  R.~Terada$^a$, Y.~Hasegawa, T.~Takeshita,   }\\[5pt]

{\large \it Department of Physics, Faculty of Science, Shinshu University \\ $^a$Graduate School of Science and Technology, Shinshu University}\\ 

\end{center}

\vspace{10pt}

\abstract{
We performed a Geant4 simulation study on showers generated by electrons and hadrons in a large homogeneous calorimeter. We found that the energy deposit can be expressed as a linear function of the track length. The line does not pass through the origin, and the energy deposit at the intercept is proportional to the incident energy. Moreover, for both electrons and hadrons, the slope of the line is independent of the incident energy. The energy resolution of the calorimeter can be expressed in terms of the distribution around the correlation line, which we found to be very good at about  $ 19\% / \sqrt{E(\rm{GeV})}$ for pions.
}

\newpage

\section{Introduction}
In high energy particle experiments, the role of the calorimeter becomes increasingly important at higher energies because energy resolution improves with energy. On the other hand, the momentum resolution of the tracker deteriorates with higher energies.
Since there is a substantial difference between an electromagnetic shower and a hadronic shower, calorimeters are generally divided into two specialized types.
One type (ECAL) is used for electromagnetic showers, while the other (HCAL) is for hadronic showers.
An ECAL is made of a high-Z material, such as lead or tungsten, with a short radiation length, while the choice of material for an HCAL depends on the nuclear interaction length.

It is important to note that hadronic showers also contain electromagnetic showers generated by neutral pions.
This phenomenon makes precise hadron energy measurements difficult. However, the energy resolution can be improved by taking complete information from both types of showers into account.
From this viewpoint, a homogeneous calorimeter is more favorable than a sampling calorimeter. In addition, a homogeneous calorimeter does not require to be separated into two parts.
Therefore, we used Geant4\cite{bib:geant4_1}\cite{bib:geant4_2} to simulate electromagnetic and hadronic showers in large, simplified, homogeneous calorimeter consisting of a PbWO$_4$ crystal\cite{bib:pbwo4_1}.
We expected excellent energy resolution for both electrons and pions from such a calorimeter by using two observables, and we found this to be the case.
The observables used were the energy deposition in the material by charged particles, which is measured from the scintillation light emission, and the charged particle track length obtained from the Cherenkov radiation.

\section{Simulation setup}
We employed Geant4 Version 10.0p1 for our simulation study.
The specific parameters used are listed below, while the other parameters were left at their default values.
The detector consisted of a PbWO$_4$ crystal with dimensions 2 m $\times$ 2 m $\times$ 2 m = 8 m$^3$, large enough to fully contain the hadronic showers.
The particles were injected into the calorimeter perpendicularly at the central point of its surface.
We used the hadron physics list FTFP\_BERT\cite{bib:ftfp_bert_1}\cite{bib:ftfp_bert_2}\cite{bib:ftfp_bert_3}.
During the GEANT4 simulations, we measured the energy deposition (ED) and the track length (TL) of the charged particles in showers created by incident particles passing through the detector material until all charged particles in the showers stopped. Afterwards, the total ED and TL values were calculated for each event.
Note that the ED and TL values were recorded within 100~ns after particle injection
into the detector in order to exclude delayed reactions by slow neutrons.
We generated 1,000 events for each energy and particle type.

\section{Results}
\begin{figure}[htbp]
  \centering
   \includegraphics[width=80mm]{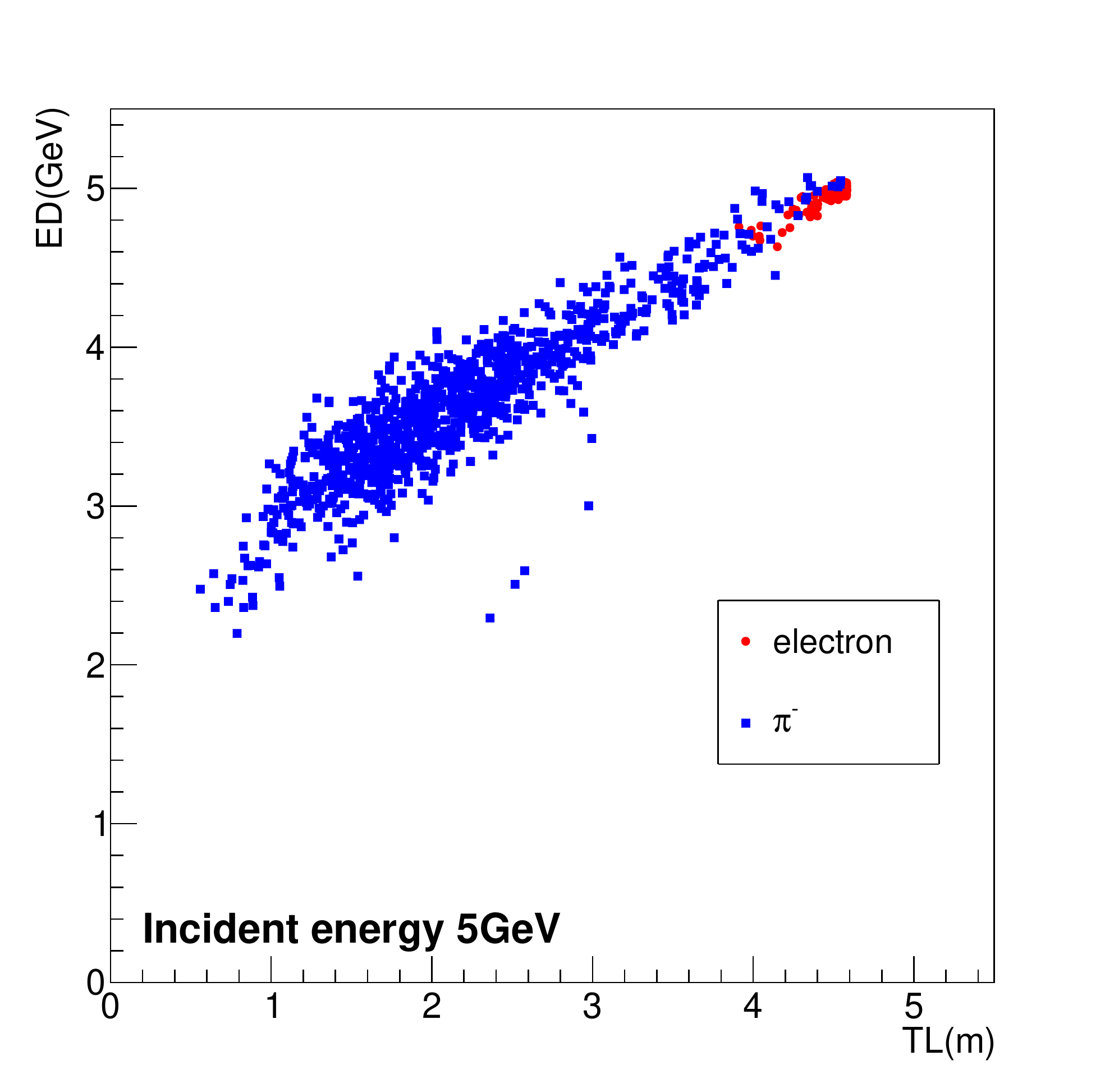}
  \caption{A correlation plot of ED and TL for 5GeV incident energy. The vertical axis is the energy deposition (ED) in GeV, and the horizontal axis is the track length (TL) in the meter for 5~GeV incident particles. Red dots represent electrons, and blue dots represent negative pions. It turns out that the correlation between ED and TL is the same for electrons as well as negative pions.}
  \label{fig:ed-td-pion-electron}
\end{figure}

Fig.~\ref{fig:ed-td-pion-electron} shows a correlation between the total ED and TL values when negative pions and electrons with 5~GeV energies were injected. 

\begin{figure}[htb]
  \centering
   \includegraphics[width=80mm]{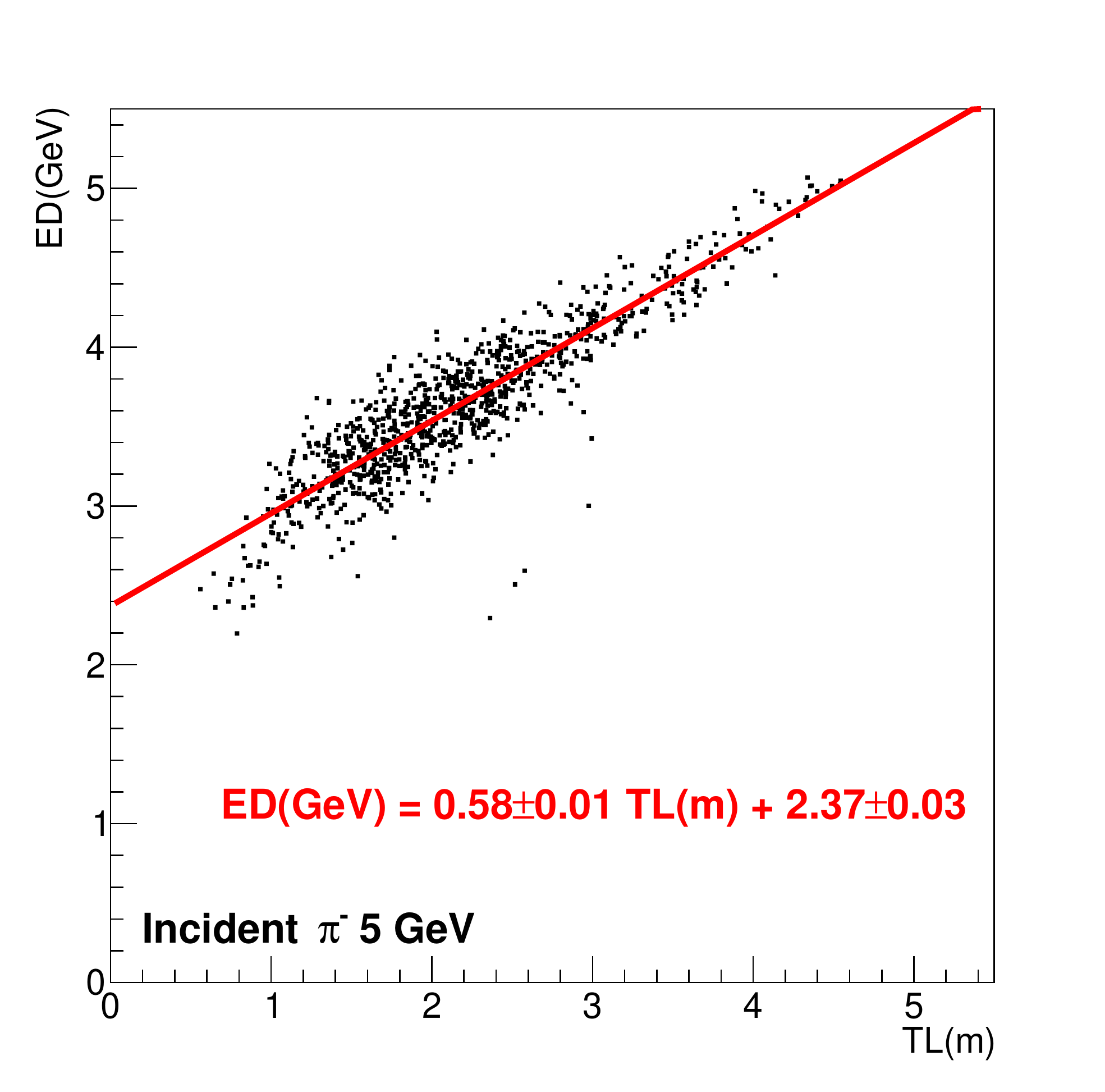}
  \caption{A correlation plot of ED and TL for 5GeV incident energy. The vertical axis is the energy deposition (ED) in GeV, and the horizontal axis is the track length (TL) in the meter for 5~GeV incident pions. The red line represents a linear fit for all points. The fit results show that the intercept is nonzero.}
  \label{fig:ed-td-pion-fit}
\end{figure}

A clear linear correlation between these observables is apparent: both pion as well as electron events occurred along the same line. We fitted the events for each incident energy and particle type with a line and determined the y-intercept and the slope, which were used to characterize the events.
Fig.~\ref{fig:ed-td-pion-fit} shows the 5~GeV negative pion events fitted to a straight line as an example.

\begin{figure}[htb]
  \begin{minipage}[h]{0.5\textwidth}
\centering
\includegraphics[width=60mm]{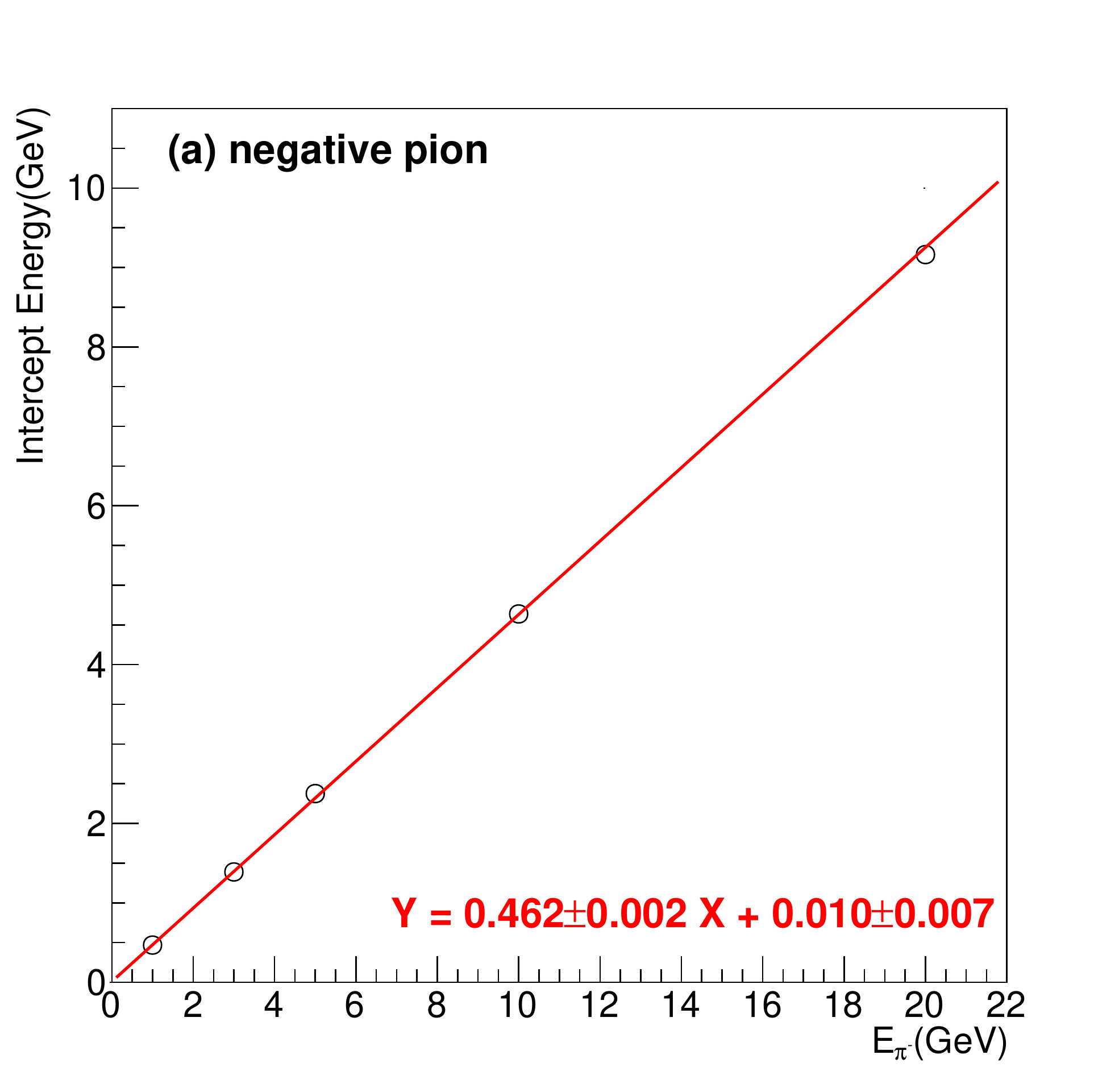}
\phantomsubcaption\label{fig:03}
\end{minipage}
\begin{minipage}[h]{0.5\textwidth}
\includegraphics[width=60mm]{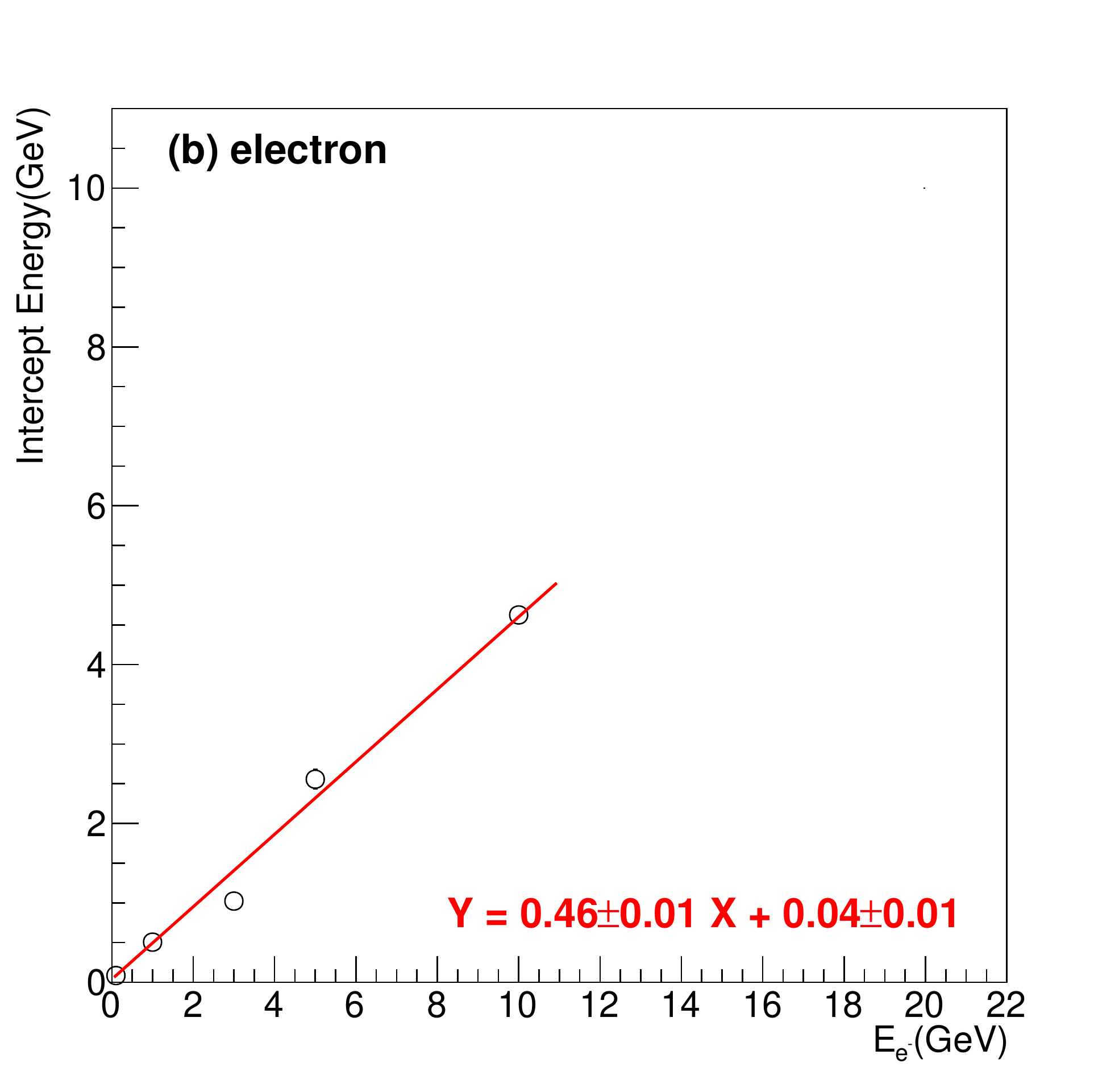}
\phantomsubcaption\label{fig:04}
\end{minipage}
\captionsetup{format=plain, margin=50pt, name=Fig}
\caption[]{Intercept energies as functions of incident energies for
  \subref{fig:03} negative pions and \subref{fig:04} electrons.
Both results confirm that the correlation between the intercept and the energy is linear.}
\label{fig:intercept}
\end{figure}

Note that the fitted line does not pass through the origin and has a positive intercept.
We found that the intercept was linearly dependent on the incident energy, as
shown in  Fig.~\ref{fig:intercept}\subref{fig:03} for pions and Fig.~\ref{fig:intercept}\subref{fig:04} for electrons.
The energy dependencies of the intercepts for electrons and pions were very similar, indicating a similar calorimeter response to both pions and electrons.

\begin{figure}[htbp]
\begin{minipage}[h]{0.5\textwidth}
\centering
   \includegraphics[width=60mm]{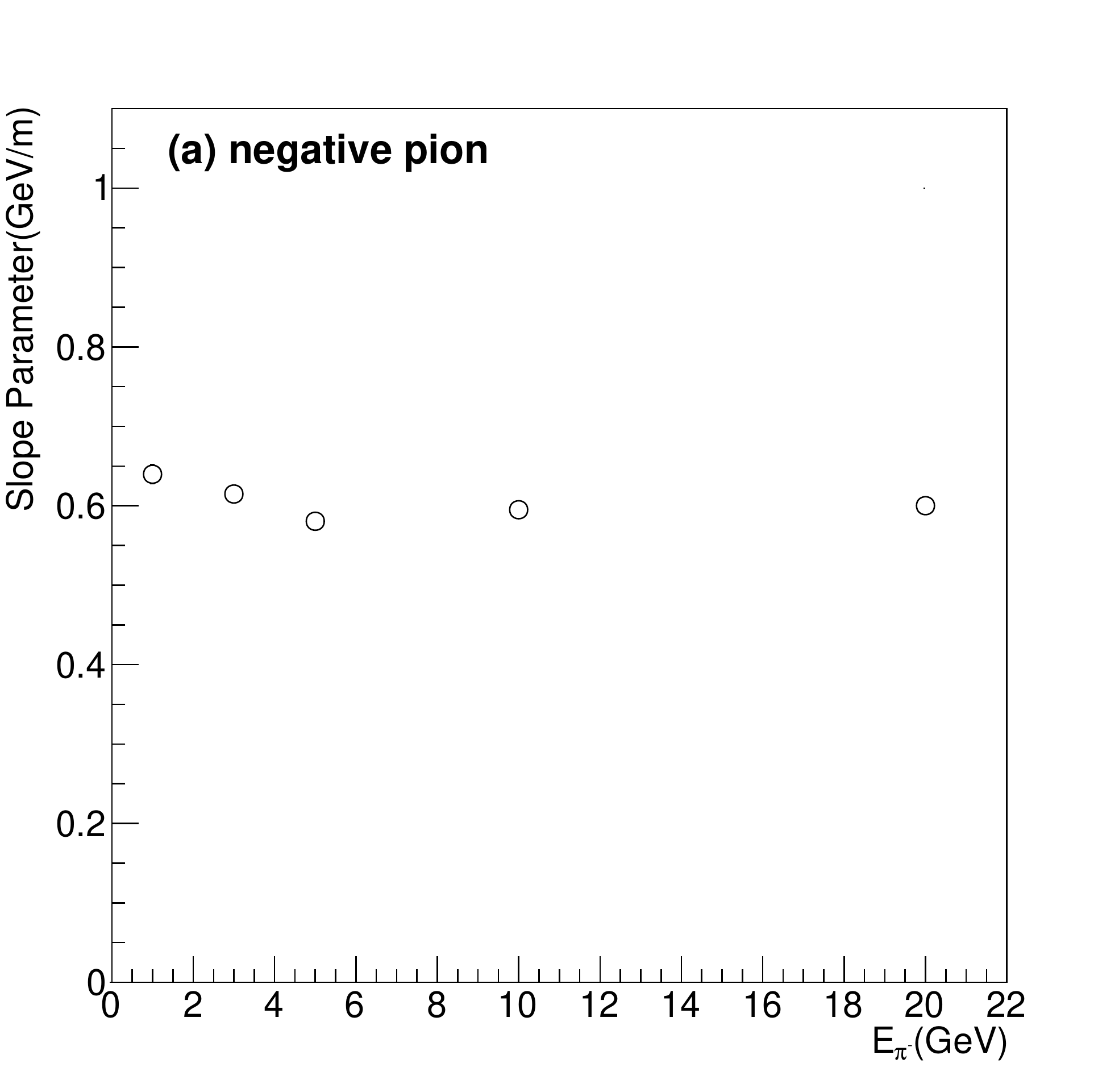}
\phantomsubcaption\label{fig:05}
\end{minipage}
\begin{minipage}[h]{0.5\textwidth}
  \centering
   \includegraphics[width=60mm]{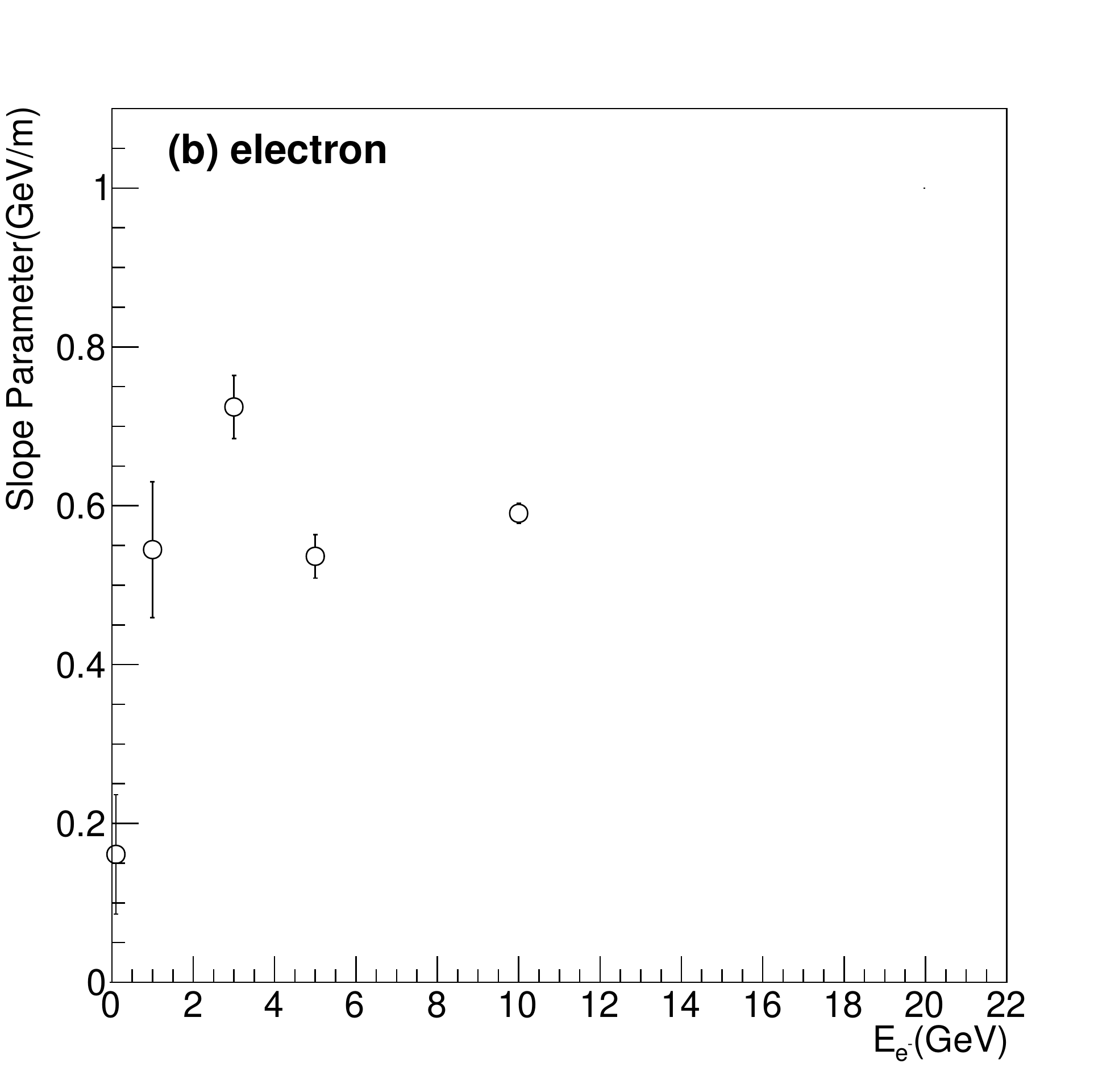}
\phantomsubcaption\label{fig:06}
\end{minipage}
\captionsetup{format=plain, margin=50pt, name=Fig}
\caption[]{Slope parameters as functions of incident energies for \subref{fig:05} negative pions and \subref{fig:06} electrons. The parameters for both types of particles have the same value independent of incident energy.}
  \label{fig:slope}
\end{figure}

Furthermore, we found that the slopes for negative pions and electrons were independent of incident energy. A similar value of about 0.6~GeV/m was obtained for both particles, as shown in Figs.~\ref{fig:slope}\subref{fig:05} and \ref{fig:slope}\subref{fig:06} for pions and electrons, respectively.
Note that the data point for electrons with $E_{e^-}<$ 100 MeV in Fig.~\ref{fig:slope}\subref{fig:06} is well below 0.6~GeV/m since the electromagnetic shower was not sufficiently developed for the energy.

\begin{figure}[htb]
  \centering
   \includegraphics[width=80mm]{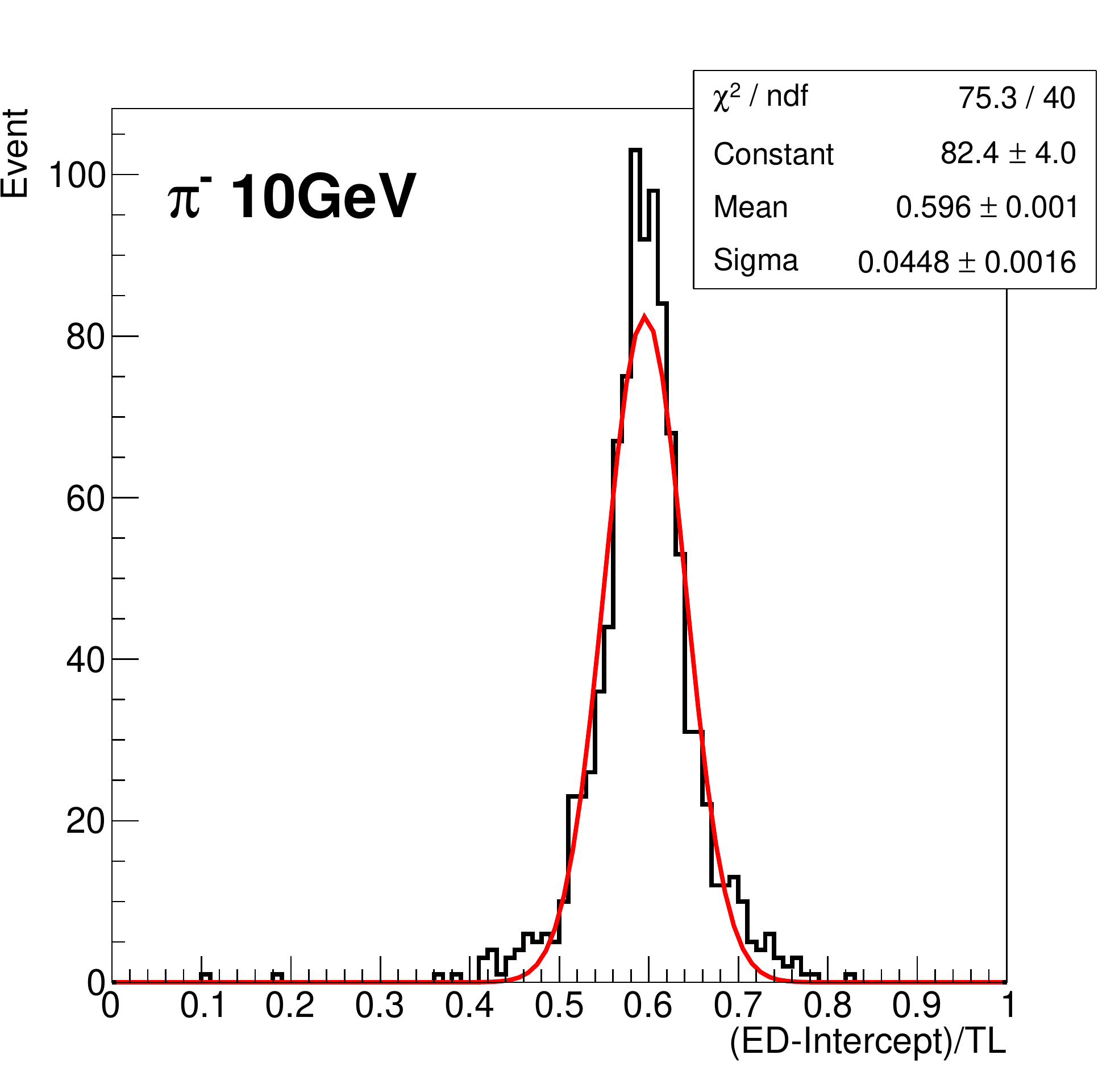}
  \caption{The distribution of the ED/TL ratio for 10~GeV negative pions. A Gaussian distribution provides a good fit.}
  \label{fig:dist-ed-td-pion}
\end{figure}

\begin{figure}[htb]
\begin{minipage}[h]{0.5\textwidth}
\centering
   \includegraphics[width=60mm]{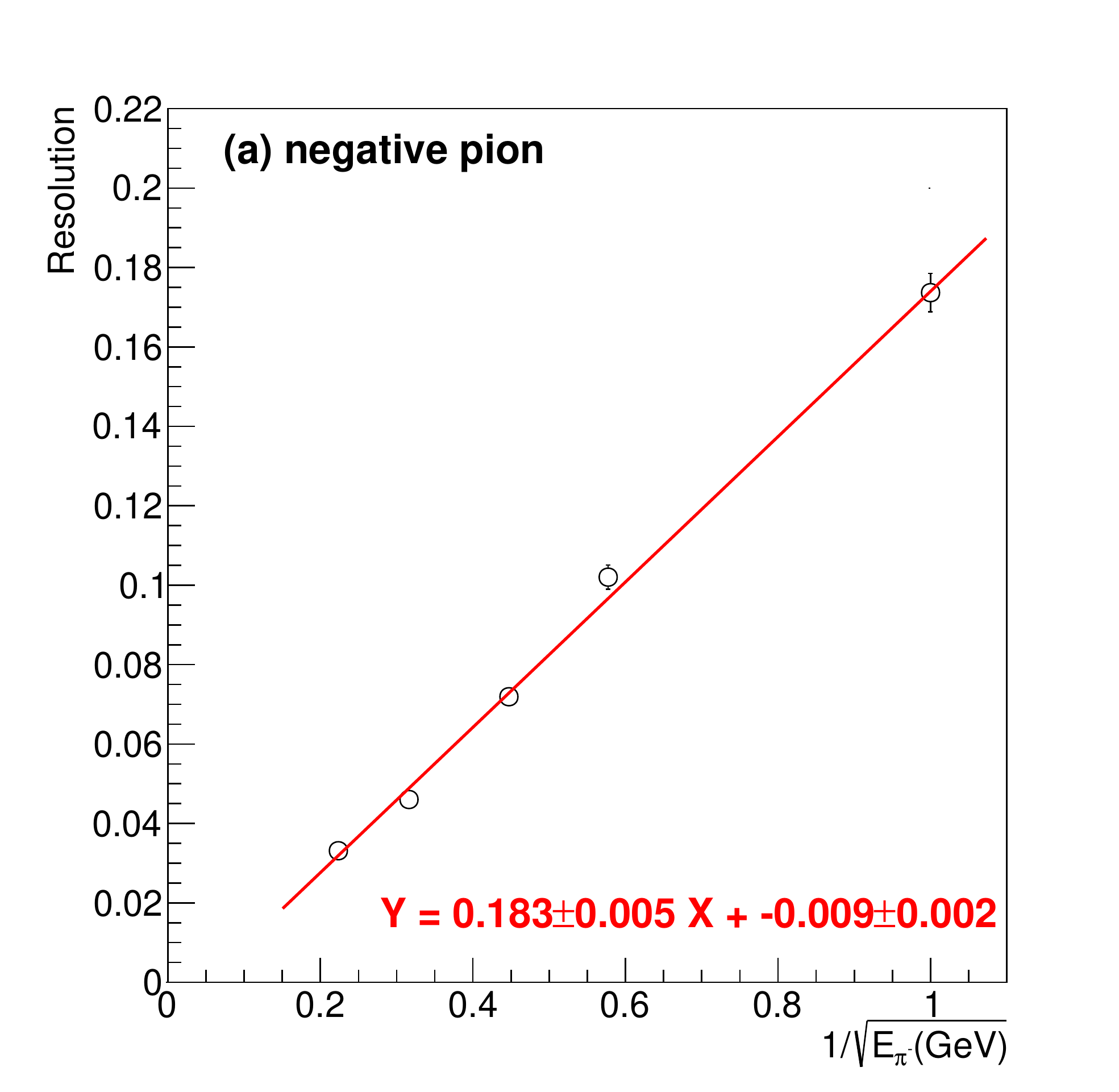}
\phantomsubcaption\label{fig:08}
\end{minipage}
\begin{minipage}[h]{0.5\textwidth}
   \includegraphics[width=60mm]{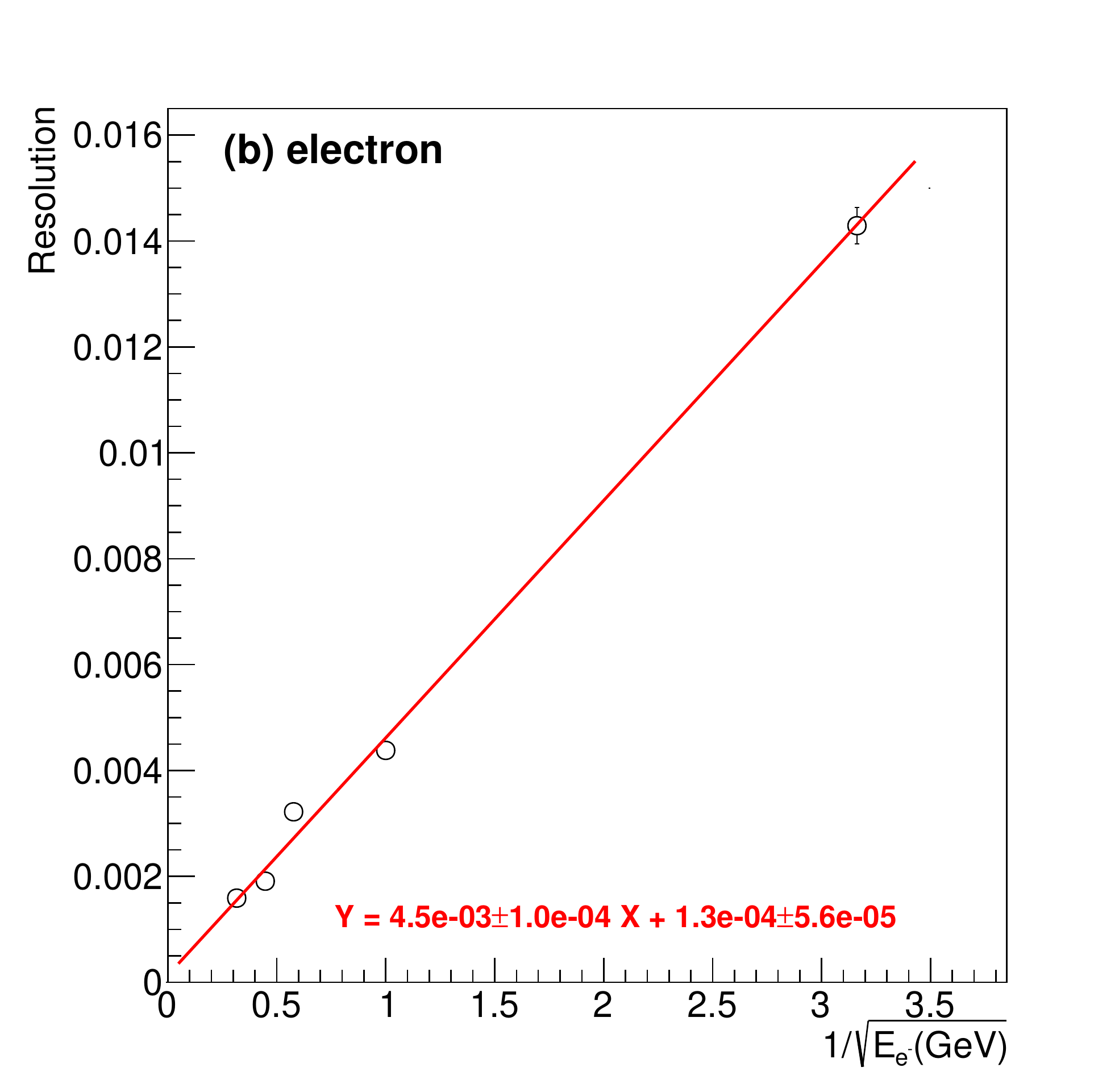}
\phantomsubcaption\label{fig:09}
\end{minipage}
\captionsetup{format=plain, margin=50pt, name=Fig}
\caption[]{Energy resolutions as functions of incident energies for
  \subref{fig:08} negative pions and \subref{fig:09} electrons. They were fitted with $1/\sqrt{E({\rm GeV})}$ and found to be 19\% and 0.5\% for pions and electrons, respectively.}
  \label{fig:resolution}
\end{figure}

Calorimeter performance is characterized by two parameters: its linearity and energy resolution.
The linearity of the calorimeter in this study is represented by the energy intercept obtained by fitting ED with a linear function of TL.
The energy resolution of the calorimeter can be expressed in terms of the spread in the measured energy. Accordingly, in Fig.~\ref{fig:dist-ed-td-pion} we plot the distribution of the ratio of ED, from which the intercept energy has been subtracted, to TL for
10~GeV negative pions. The results match a Gaussian distribution very well.
The energy resolution can thus be expressed in terms of the width of the ratio distribution.
Figs.~\ref{fig:resolution}\subref{fig:08} and \subref{fig:09} show the energy resolutions as functions of $1/\sqrt{E}$ for pions and electrons, respectively. When the resolution was parametrized as $\sigma_E / E = ({\rm stochastic~term}) /\sqrt{E({\rm GeV})} + (constant~term)$, we found that the stochastic terms for pions and  electrons were 19\% and 0.5\%, respectively.

\begin{figure}[htb]
  \centering
   \includegraphics[width=80mm]{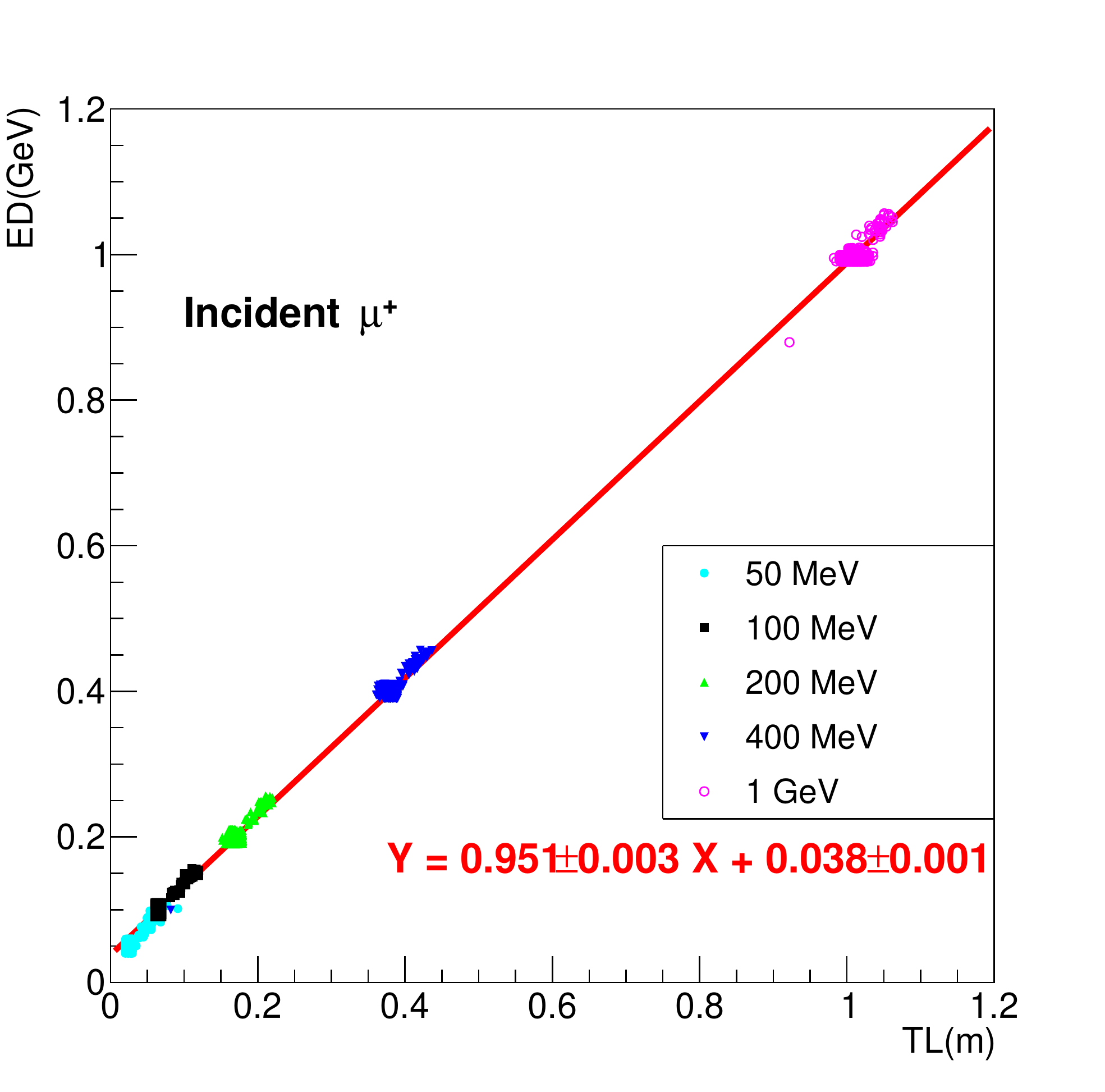}
  \caption{ED versus TD for anti-muons with kinetic energies of 50, 100,
    200, 400, and 1000 MeV. The red line is the result of a linear fit
    to all points, and its intercept is about 38 MeV (not zero).}
  \label{fig:ed-td-muon}
\end{figure}

It is important that correlation between ED and TL result has a non-zero intercept.
We examined the effect of the Bragg curve on anti-muons, which do not create showers, with kinetic energies of 50, 100, 200, 400, and 1000 MeV before they decayed, as shown in Fig.~\ref{fig:ed-td-muon}.
Using these data, we found the value of intercepts to be independent of muon's kinetic energy.
The value determined from the ED versus TL plot was about 38~MeV at zero kinetic energy limit.
This intercept could be understood by the Bragg curve effect.

\begin{figure}[htb]
  \centering
   \includegraphics[width=80mm]{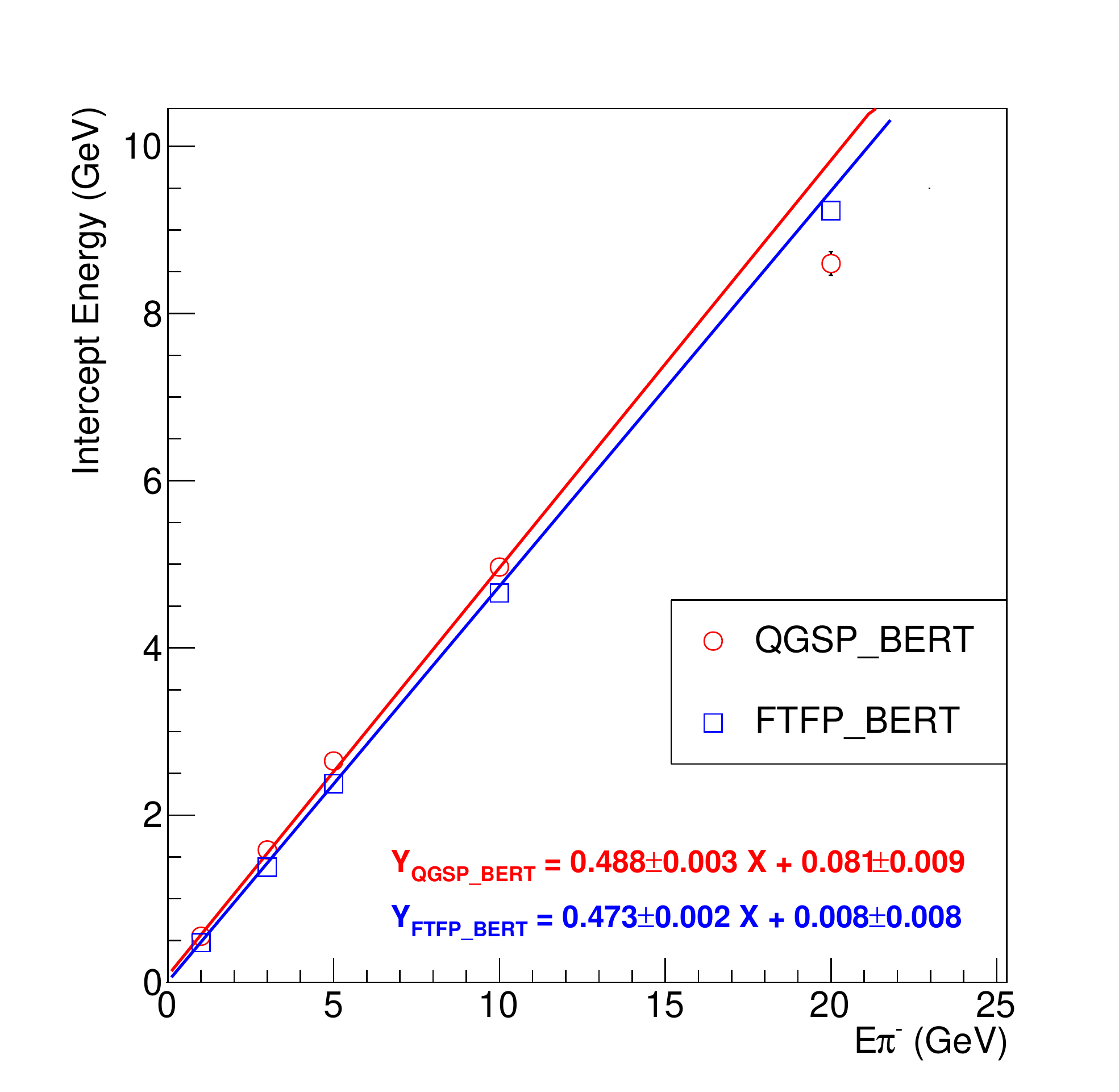}
  \caption{Comparison of line intercepts between different hadron lists (QGSP\_BERT and FTFP\_BERT).
  Linear fitting was performed between 1 to 10~GeV.
  The difference in hadron physics lists results in a systematic difference between the intercepts.
  QGSP\_BERT switches between BERT and QGSP at around 10~GeV.}
  \label{fig:gfsp-intercept}
\end{figure}

\begin{figure}[htb]
  \centering
   \includegraphics[width=80mm]{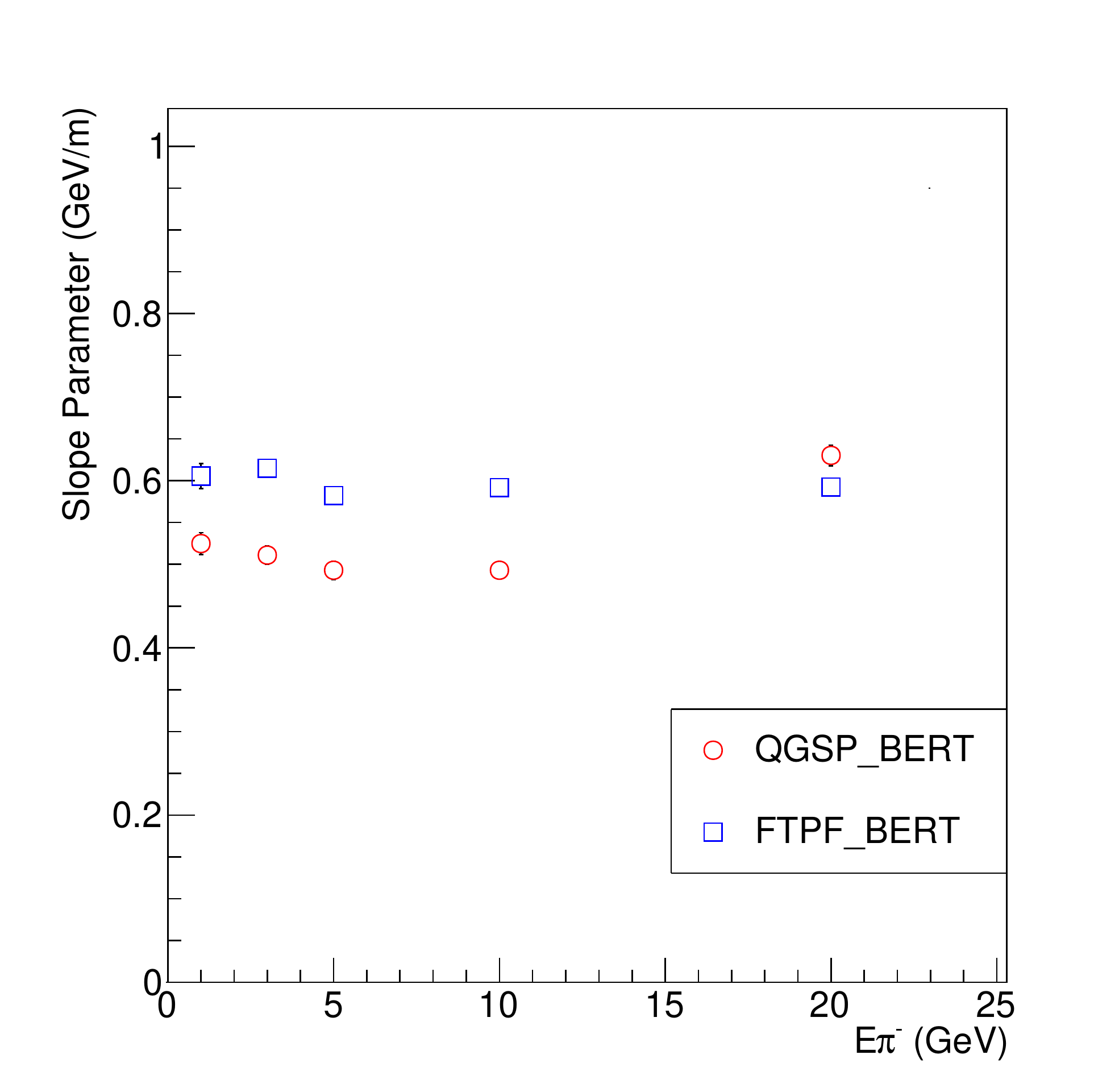}
  \caption{Comparison of slope parameters between different hadron lists (QGSP\_BERT and FTFP\_BERT).
  The difference in hadron physics lists results in a systematic difference in the slope parameters.
  QGSP\_BERT switches between BERT and QGSP at around 10~GeV resulting in a significantly higher slope parameter for the 20~GeV incident energy.}
  \label{fig:gfsp-slope}
\end{figure}

The results so far were obtained using a hadron physics list called FTFP\_BERT. 
There exits several other lists that reflect the differences in phenomenological hadron interpretations.
Therefore, another simulation was performed using QGSP\_BERT instead of FTFP\_BERT in order to check for systematic uncertainties due to difference in hadronic process modeling.
Figs.~\ref{fig:gfsp-intercept} and \ref{fig:gfsp-slope} show the dependencies of intercept energies and slope parameters on incident energies between the two hadron models, respectively.
The difference in the hadron lists resulted in a systematic difference in both intercept as well as slope parameters.
Moreover, the 20~GeV QGSP\_BERT point deviates from the fitted line
because the boundary between the energy regions is located at around 10~GeV\cite{bib:physics_model_diff}. BERT is used for energies below this boundary, whereas QGSP is used for energies above it. On the other hand, for FTFP\_BERT the boundary between BERT and FTFP regions is around 5~GeV resulting in a small deviation from the fitted line.

\clearpage

\section{Discussion}

We showed that a linear fit is an excellent representation of the correlation between ED and TL for a homogeneous calorimeter.
We also investigated the reason for the constant offset in ED,
which is represented by the intercept of ED as a linear function of TL, and we found that it can be understood as a consequence of the Bragg curve.
As a charged particle slows down, the ED increases more significantly than it does at high energies.
Near the endpoint of the Bragg curve, the ED deposition is large, but the TL does not change.
However, in showers many particles contribute to the constant term and, the Bragg curve influence appears in every track.
Therefore, as shower energy increases, the constant term in the ED-TL relation increases linearly since the number of particles in the shower increases.

The validity of this study was demonstrated by comparing the results from FTFP\_BERT with those using different hadron lists. Specifically, the QGSP\_BERT list was chosen as it differs from FTFP\_BERT in the high energy region. In this simulation, we also found that there was a linear relationship between ED and TL, the slope was constant regardless of the incident energy, and the intercept of the straight line was directly proportional to the incident energy. Therefore it was confirmed that our results are applicable for different hadron models.

\clearpage

\section{Conclusion}

We examined the relationship between ED and TL in a homogeneous, high-energy calorimeter using Geant4. Our simulation shows that it has good linearity and excellent energy resolution.
For a PbWO$_4$ calorimeter specifically, we found an incredible hadron resolution of about $19\% /\sqrt{E({\rm GeV})}$. We plan to verify these results experimentally: ED can be determined from the amount of scintillation emission, while TL can be obtained from the corresponding amount of Cherenkov radiation in the PbWO$_4$ crystal.

\end{document}